\def\papertitle{Learning Nonlinear Dynamics in Physical Modelling Synthesis using Neural Ordinary Differential Equations}
\def\paperauthorA{Victor Zheleznov}
\def\paperauthorB{Stefan Bilbao}
\def\paperauthorC{Alec Wright}
\def\paperauthorD{Simon King}
\documentclass[twoside,a4paper]{article}
\usepackage{etoolbox}
\usepackage{dafx_25}
\usepackage{amsmath,amssymb,amsfonts,amsthm}
\usepackage{euscript}
\usepackage[T1]{fontenc}
\usepackage[utf8]{inputenc}
\usepackage{ifpdf}
\usepackage[english]{babel}
\usepackage{caption}
\usepackage{subfig} % or can use subcaption package
\usepackage{color}

\input glyphtounicode
\ninept

\newcounter{numauth}
\newcounter{listcnt}
\newcommand\authcnt[1]{\ifdefined#1 \stepcounter{numauth} \fi}

\newcommand\addauth[1]{
\ifdefined#1 
\stepcounter{listcnt}
\ifnum \value{listcnt}<\value{numauth}
\appto\authorslist{, #1}
\else
\appto\authorslist{~and~#1}
\fi
\fi}
\authcnt{\paperauthorB}
\authcnt{\paperauthorC}
\authcnt{\paperauthorD}
\authcnt{\paperauthorE}
\authcnt{\paperauthorF}
\authcnt{\paperauthorG}
\authcnt{\paperauthorH}
\authcnt{\paperauthorI}
\authcnt{\paperauthorJ}
\def\authorslist{\paperauthorA}
\addauth{\paperauthorB}
\addauth{\paperauthorC}
\addauth{\paperauthorD}
\addauth{\paperauthorE}
\addauth{\paperauthorF}
\addauth{\paperauthorG}
\addauth{\paperauthorH}
\addauth{\paperauthorI}
\addauth{\paperauthorJ}
\usepackage{times}
\newif\ifpdf
\ifx\pdfoutput\relax
\else
   \ifcase\pdfoutput
      \pdffalse
   \else
      \pdftrue
   \fi
\fi

\ifpdf % compiling with pdflatex
  \usepackage[pdftex,
    pdftitle={\papertitle},
    pdfauthor={\authorslist},
    pdfsubject={Proceedings of the 28th International Conference on Digital Audio Effects (DAFx25)},
    colorlinks=false, % links are activated as color boxes instead of color text
    bookmarksnumbered, % use section numbers with bookmarks
    pdfstartview=XYZ % start with zoom=100% instead of full screen; especially useful if working with a big screen :-)
  ]{hyperref}
  \usepackage[pdftex]{graphicx}
\else % compiling with latex
  \usepackage[dvips]{epsfig,graphicx}
  \usepackage[dvips,
    pdftitle={\papertitle},
    pdfauthor={\authorslist},
    pdfsubject={Proceedings of the 28th International Conference on Digital Audio Effects (DAFx25)},
    colorlinks=false, % no color links
    bookmarksnumbered, % use section numbers with bookmarks
    pdfstartview=XYZ % start with zoom=100% instead of full screen
  ]{hyperref}
\fi
\usepackage[hypcap=true]{caption}
\title{\papertitle}

 \twoaffiliations{
 \paperauthorA\,\sthanks{This work was supported by the SGSAH AHRC Doctoral Training Partnership [grant number AH/R012717/1]; and the University of Edinburgh.}, \paperauthorB, \paperauthorC
 }
 {\href{https://www.acoustics.ed.ac.uk/}{Acoustics and Audio Group}\\ University of Edinburgh\\ Edinburgh, UK\\
 {\tt \{\href{mailto:v.zheleznov@ed.ac.uk}{v.zheleznov}, \href{mailto:s.bilbao@ed.ac.uk}{s.bilbao}, \href{mailto:alec.wright@ed.ac.uk}{alec.wright}\}@ed.ac.uk}
 }
 {\paperauthorD}
 {\href{https://www.cstr.ed.ac.uk/}{Centre for Speech Technology Research}\\ University of Edinburgh\\ Edinburgh, UK\\
 {\tt \href{mailto:simon.king@ed.ac.uk}{simon.king@ed.ac.uk}}
 }

\usepackage{bm}
\usepackage{xcolor}

\usepackage{siunitx}
\newcommand{\numberthis}{\addtocounter{equation}{1}\tag{\theequation}}

\newcommand{\norm}[1]{\left\lVert#1\right\rVert}
\newcommand{\module}[1]{\left\vert#1\right\vert}
\newcommand*{\tran}{^\mathrm{T}}

\usepackage{mathtools} % for \mathclap
\usepackage{comment}
\begin{document}
% more pdf-tex settings:
\ifpdf % used graphic file format for pdflatex
  \DeclareGraphicsExtensions{.png,.jpg,.pdf}
\else  % used graphic file format for latex
  \DeclareGraphicsExtensions{.eps}
\fi

%\makeatletter
%\pdfbookmark[0]{\@pdftitle}{title}
%\makeatother

\maketitle

\begin{abstract}
Modal synthesis methods are a long-standing approach for modelling distributed musical systems. In some cases extensions are possible in order to handle geometric nonlinearities. One such case is the high-amplitude vibration of a string, where geometric nonlinear effects lead to perceptually important effects including pitch glides and a dependence of brightness on striking amplitude. A modal decomposition leads to a coupled nonlinear system of ordinary differential equations. Recent work in applied machine learning approaches (in particular neural ordinary differential equations) has been used to model lumped dynamic systems such as electronic circuits automatically from data. In this work, we examine how modal decomposition can be combined with neural ordinary differential equations for modelling distributed musical systems. The proposed model leverages the analytical solution for linear vibration of system's modes and employs a neural network to account for nonlinear dynamic behaviour. Physical parameters of a system remain easily accessible after the training without the need for a parameter encoder in the network architecture. As an initial proof of concept, we generate synthetic data for a nonlinear transverse string and show that the model can be trained to reproduce the nonlinear dynamics of the system. Sound examples are presented.
\end{abstract}

%%%%%%%%%% INTRODUCTION %%%%%%%%%%
\section{Introduction}

Research into physical modelling synthesis has a long history. Various simulation techniques have been employed, including finite-difference time-domain methods \cite{BilbaoNSS}, modal synthesis \cite{Morrison1993} and port-Hamiltonian methods \cite{Falaize2017}. These approaches all rely on the numerical solution of a set of ordinary or partial differential equations (ODEs/PDEs) that describe the dynamics of a system and are accompanied by initial and boundary conditions, and external forces or input signals. In contrast, machine learning approaches generally construct musical systems automatically from data, often in a black-box manner, and recently have gained popularity in audio research, especially in virtual-analog modelling \cite{Wright2019}. Some data-driven approaches --- such as neural ordinary differential equations (NODEs), in which a derivative of system's state vector is parameterised by a neural network \cite{Chen2018} --- incorporate existing knowledge about the underlying system into their training \cite{Raissi2019}. As electronic circuits can generally be viewed as finite-dimensional systems, they are well-modelled using NODEs \cite{Wilczek2022}.

However, modelling of distributed musical systems such as strings, plates, etc.\@ using machine learning approaches has seen limited attention in the literature. Parker et al.\@ \cite{Parker2022} have presented recurrent neural networks for physical modelling based on a Fourier neural operator \cite{Li2021}. These recurrent structures are trained from data spanning a few initial time steps of around $2$ \unit{\milli\second}. Extrapolation in time is tested over intervals spanning up to 10 times that seen during training and the best performing model shows degradation towards the end of the simulation. Diaz et al.\@ \cite{Diaz2024} have introduced a Koopman-based model that addresses some discrepancies compared with recurrent architectures such as solution accuracy but extrapolation in time remains a challenge. Lee et al.\@ \cite{Lee2024} have employed differentiable digital signal processing techniques for the simulation of a nonlinear string. Amplitudes and frequencies of oscillators, corresponding to modes of a string, are adjusted using multilayer perceptron (MLP) blocks, which are trained to capture nonlinear effects.

One common drawback of these approaches is that initial conditions and physical parameters of a system, affecting pitch, timbre and other sonic characteristics, can not be modified after training, or the network architecture requires a parameter encoder to condition the solution, leading to more trainable parameters and the requirement of a larger dataset containing ground truth data for desired configurations of a system. In addition, system excitation follows from different choices of initial conditions which does not correspond to a realistic playing scenario, where external forcing terms are always present \cite{BilbaoNSS}. In this work, we aim to more tightly integrate the physics of a distributed system into a machine learning framework. In particular, we use modal decomposition to construct a system of finite dimension and separate the linear and nonlinear parts of the problem. Then, we replace only a dimensionless memoryless nonlinearity (that describes coupling between the modes) with a neural network and train NODEs to obtain the resulting model. Consequently, physical parameters remain easily accessible and the model generalises to physical parameters, sampling rates and time scales not seen during training. Compared to a fully physical model, we show that the proposed approach is more computationally efficient for the case of nonlinear transverse string vibration.

The paper is organised as follows. A simple model of nonlinear transverse string vibration is described in Section \ref{sec: string}. Section \ref{sec: modal} derives modal equations for a string, which are discretised in time for computer simulation in Section \ref{sec: discretisation}. Section \ref{sec: algorithm} outlines the proposed learning algorithm, which is evaluated in Section \ref{sec: evaluation} across several case studies. Sound examples are available on the accompanying page\footnote{\url{https://victorzheleznov.github.io/dafx25}}.

%%%%%%%%%% NONLINEAR TRANSVERSE STRING %%%%%%%%%%
\section{Nonlinear Transverse String Vibration}
\label{sec: string}

The general equation of motion describing a transverse vibration of a nonlinear string in a single polarisation is:
\[
\mathcal{L} u
=
\mathcal{F}
+
\mathcal{F}_{\mathrm{e}}.
\label{eq: string}
\numberthis
\]
Here $u = u(x, t)\!\!: [0, L] \times \mathbb{R}^+ \rightarrow \mathbb{R}$ denotes the transverse displacement of a string of length $L$ and depends on spatial coordinate $x$ in \unit{\meter} and time $t$ in \unit{\sec}. Initial conditions are assumed to be zero. The string is assumed to be simply supported at both ends, implying the following boundary conditions:
\[
u(0, t)
=
\partial_x^2
u(0, t)
=
u(L, t)
=
\partial_x^2
u(L, t)
=
0,
\quad
\forall
t
\in
\mathbb{R}^+
.
\]
Output is assumed to be drawn directly from the string displacement at position $x_{\mathrm{o}}$ as $w(t) = u(x_{\mathrm{o}}, t)$.

%%%%%%%%%% LINEAR VIBRATION %%%%%%%%%%
\subsection{Linear Vibration}

The linear part of the string vibration is encapsulated in the operator $\mathcal{L}$, defined as:
\[
\mathcal{L}
=
\rho A
\partial_{t}^2
-
T \partial_{x}^2
+
EI \partial_{x}^4
+
2 \sigma_0 
\rho A
\partial_t
-
2 \sigma_1
\rho A
\partial_t\partial_{x}^2,
\label{eq: linear_term}\numberthis
\]
where $\partial_t$ and $\partial_x$ represent partial derivatives with respect to $x$ and $t$, respectively. Physical parameters that appear in \eqref{eq: linear_term} are: the material density $\rho$ in \unit{\kilogram\meter\tothe{-3}}; the string cross-sectional area $A = \pi r^2$ in \unit{\meter\tothe{2}} for a string of radius $r$; the tension $T$ in \unit{\newton}; Young's modulus $E$ in \unit{\newton\per\meter\tothe{2}}; and moment of inertia $I = \frac{1}{4} \pi r^4$ in \unit{\meter\tothe{4}}. Frequency-independent and dependent loss is characterised by parameters $\sigma_0 \geq 0$ and $\sigma_1 \geq 0$, respectively. See \cite{BilbaoNSS} for more on this term in the context of linear string vibration.

%%%%%%%%%% NONLINEARITY %%%%%%%%%%
\subsection{Nonlinearity}

Nonlinear dynamics of the string are described in a force density $\mathcal{F}$. A general model for $\mathcal{F}$ is given by Morse and Ingard \cite{MorseAcoustics} and includes both longitudinal and transverse motion of a string in two polarisations. In this work we neglect the longitudinal motion and one of the two polarisations, leading to the following force density after taking a Taylor series expansion on the force potential \cite{Bilbao2005}:
\[
\mathcal{F}
(x, t)
=
\partial_x
\bigg(
\frac{EA - T}{2}
\xi^3
\bigg),
\quad
\xi
\triangleq
\partial_x u.
\label{eq: nonlinearity}\numberthis
\]
For the force density \eqref{eq: nonlinearity} to be conservative we assume $EA \geq T$, which is true in the case of musical strings. Compared to the Kirchhoff-Carrier model \cite{KirchhoffVorlesungen, Carrier1945}, which adequately reproduces only the pitch glide effect, model \eqref{eq: nonlinearity} is capable of capturing other perceptually important effects such as phantom partials \cite{BilbaoNSS}.

%%%%%%%%%% PLUCKING EXCITATION %%%%%%%%%%
\subsection{Plucking Excitation}

The string is excited by a pointwise external force $\mathcal{F}_{\mathrm{e}}$, which can be modelled as:
\[
\mathcal{F}_{\mathrm{e}}
(x, t)
=
\delta(x - x_{\mathrm{e}})
f_{\mathrm{e}}(t)
,
\]
where $\delta(x - x_{\mathrm{e}})$ is the Dirac delta function at the excitation position $x_{\mathrm{e}}$. Function $f_{\mathrm{e}}(t)$ resembles a pluck of a string and is of the following form \cite{Bilbao2019}:
\[
f_{\mathrm{e}}(t)
=
\begin{cases}
\frac{1}{2}f_{\mathrm{amp}}
\big[
1
-
\cos
\big(
\frac{\pi t}{T_{\mathrm{e}}}
\big)
\big]
,
\quad
&t
\in
[
0,
T_{\mathrm{e}}
]
\\
0,
&\text{otherwise}
\end{cases}
\label{eq: excitation}
\numberthis
\]
Here $f_{\mathrm{amp}}$ is the excitation amplitude in \unit{\newton} and $T_{\mathrm{e}}$ is the excitation duration in \unit{\sec}. The excitation starting time is assumed to be zero.

%%%%%%%%%% EQUATION SCALING %%%%%%%%%%
\subsection{Equation Scaling}

In view of using the string model \eqref{eq: string} for dataset generation, it is useful to reduce the number of physical parameters to the smallest possible set. Firstly, we employ spatial scaling by introducing a normalised position variable $x'=\frac{x}{L} \in [0,1]$:
\begin{multline*}
\partial_{t}^2u
=
\gamma^2 \partial_{x'}^2u
-
\kappa^2 \partial_{x'}^4u
-
2 \sigma_0 \partial_tu
+
2 \sigma_1' \partial_t\partial_{x'}^2u
+{}\\
{}+
\gamma^2
\partial_{x'}
\bigg(
\frac{1}{2L^2}
\bigg(
\frac{EA}{T}
-
1
\bigg)
(\partial_{x'}u)^3
\bigg)
+
\frac{1}{\rho A L}
\delta(x' - x_{\mathrm{e}}')
f_{\mathrm{e}}(t),
\end{multline*}
where 
$\gamma = \frac{1}{L}\sqrt{\frac{T}{\rho A}}$,
$\kappa = \frac{1}{L^2}\sqrt{\frac{EI}{\rho A}}$
and
$\sigma_1' = \frac{\sigma_1}{L^2}$.

Secondly, we scale the displacement by $u' = u\cdot u_0$ and obtain:
\begin{multline*}
\partial_{t}^2u'
=
\gamma^2 \partial_{x'}^2u'
-
\kappa^2 \partial_{x'}^4u'
-
2 \sigma_0 \partial_tu'
+
2 \sigma_1' \partial_t\partial_{x'}^2u' 
+{} \\
{}+
\gamma^2 \partial_{x'}(\xi'^3)
+
\delta(x' - x_{\mathrm{e}}')
f_{\mathrm{e}}'(t),
\label{eq: string_scaled}
\numberthis
\end{multline*}
where
$u_0 = \frac{1}{L}\sqrt{\frac{1}{2}(\frac{EA}{T} - 1)}$,
$\xi' \triangleq \partial_{x'} u'$
and
$f_{\mathrm{e}}'(t) = \frac{u_0}{\rho A L} f_{\mathrm{e}}(t)$.

Thus, we have reduced a set of physical parameters $\{L, \rho, A,\allowbreak T, E, I, \sigma_0, \sigma_1\}$ to a set of only four parameters $\{\gamma, \kappa, \sigma_0, \sigma_1'\}$. In the following sections we omit the prime while referring to the scaled string model \eqref{eq: string_scaled}.

%%%%%%%%%% MODAL DECOMPOSITION %%%%%%%%%%
\section{Modal Decomposition}
\label{sec: modal}

The solution to equation \eqref{eq: string_scaled} can be decomposed into a set of modes, yielding a finite-dimensional system when truncated to finite order $M$. The transverse displacement $u$ is rewritten as a superposition of modal displacements $\bm {q}(t) = [q_{1}(t),\dots,q_{M}(t)]\tran$:
\[
u(x, t)
=
\sum\limits_{m=1}^{M}
\Phi_m(x) q_m(t)
=
\bm{\Phi}\tran(x)
\bm{q}(t)
.
\label{eq: superposition}
\numberthis
\]
Here modal shapes $\Phi_m(x) = \sqrt{2}\sin(m \pi x),\;m=1,\dots,M$ correspond to the solution of the eigenvalue problem for a stiff string under simply supported boundary conditions \cite{BilbaoNSS}.

Substituting \eqref{eq: superposition} into \eqref{eq: string_scaled}, left-multiplying by $\bm{\Phi}$ and taking an $L^2$ inner product over the interval $[0, 1]$, we obtain the following second-order system of ODEs:
\[
\ddot{\bm{q}}
+
2 \bm{\mathrm{S}}
\dot{\bm{q}}
+
\bm{\Omega}^2
\bm{q}
=
\gamma^2
\bm{f}(\bm{q})
+
\bm{\Phi}(x_{\mathrm{e}})
f_{\mathrm{e}}(t),
\label{eq: modal_equation}
\numberthis
\]
where $\bm{\mathrm{S}}$ and $\bm{\Omega}$ are $M \times M$ diagonal matrices, defined using modal wavenumbers $\beta_m = m\pi,\;m=1,\dots,M$ as:
\[
\mathrm{S}_{m,m}
=
\sigma_0
+
\sigma_1
\beta_m^2, \quad
\Omega_{m,m}
=
\sqrt{
\gamma^2
\beta_m^2
+
\kappa^2
\beta_m^4
}
.
\]

%%%%%%%%%% NONLINEARITY %%%%%%%%%%
\subsection{Nonlinearity}

To obtain a closed-form expression for dimensionless nonlinearity $\bm{f}(\bm{q})\!\!: \mathbb{R}^M \rightarrow \mathbb{R}^M$, we firstly rewrite the partial derivative $\xi$ in the modal form:
\[
\xi
=
\partial_x
\Bigg(
\sum_{i_1=1}^M
\Phi_{i_1} q_{i_1}
\Bigg)
=
\sqrt{2}
\pi
\sum_{i_1=1}^M
i_1
\underbrace{
\cos(i_1 \pi x)
}_{\triangleq c_{i_1}}
q_{i_1}.
\label{eq: modal_xi}
\numberthis
\]
Raising \eqref{eq: modal_xi} to the power of $3$ and taking a partial derivative with respect to $x$, we get:
\begin{multline*}
\partial_x
\big(\xi^3\big)
=
-2\sqrt{2}\pi^4
\sum_{i_1=1}^M
\sum_{i_2=1}^M
\sum_{i_3=1}^M
\Big(
i_1^2
i_2
i_3
s_{i_1}
c_{i_2}
c_{i_3}
+{} \\
{}+
i_1
i_2^2
i_3
c_{i_1}
s_{i_2}
c_{i_3}
+
i_1
i_2
i_3^2
c_{i_1}
c_{i_2}
s_{i_3}
\Big)
q_{i_1}
q_{i_2}
q_{i_3}
,
\label{eq: modal_xi_cubic_partial}
\numberthis
\end{multline*}
where $s_{i_k} \triangleq \sin(i_k \pi x),\;k = 1,2,3$. Multiplying \eqref{eq: modal_xi_cubic_partial} by $\Phi_m$ and taking an $L^2$ inner product over the interval $[0, 1]$, we find:
\begin{align*}
f_m(\bm{q})
=
\int\limits_0^1
\Phi_m
\partial_x
\big(
\xi^3
\big)
\;
dx
=
-
\sum\limits_{i_1=1}^M
\sum_{i_2=1}^M
\sum_{i_3=1}^M
A_{i_1,i_2,i_3}^m
q_{i_1}
q_{i_2}
q_{i_3},
\label{eq: nonlinearity_modal}
\numberthis
\end{align*}
where:
\begin{align*}
A_{i_1,i_2,i_3}^m
&=
\frac{\pi^4}{2} 
\Big[
B_{i_2,i_3}^{i_1,m}
+
B_{i_3,i_1}^{i_2,m}
+
B_{i_1,i_2}^{i_3,m}
\Big],
\\
B_{i,j}^{k,m}
&=
i j k^2
\big[
\phantom{{}+{}}
\delta_{k+i,m+j}
-
\delta_{k+i,-(m+j)}
+{}
\\
&\phantom{{}=i j k^2\big[}
{}+
\delta_{k+i,m-j}
-
\delta_{k+i,-(m-j)}
+{}
\\
&\phantom{{}=i j k^2\big[}
{}+
\delta_{k-i,m+j}
-
\delta_{k-i,-(m+j)}
+{}
\\
&\phantom{{}=i j k^2\big[}
{}+
\delta_{k-i,m-j}
-
\delta_{k-i,-(m-j)}
\big].
\label{eq: tensor}
\numberthis
\end{align*}
Here $\delta_{i,j}$ denotes the Kronecker delta function. Thus, the dimensionless nonlinearity $\bm{f}(\bm{q})$ is defined by a sparse tensor $A_{i_1,i_2,i_3}^m$ \eqref{eq: tensor} which is symmetric with respect to lower indices $i_1,i_2,i_3$. Depending on the derivation, one could obtain other equivalent closed-form expressions for the tensor $A_{i_1,i_2,i_3}^m$.

%%%%%%%%%% TIME DISCRETISATION %%%%%%%%%%
\section{Time Discretisation}
\label{sec: discretisation}

For this initial study, we are using the St{\"o}rmer-Verlet method \cite{Hairer2003} as it is an explicit and efficient numerical method, although does not guarantee stability. We choose a time step $k$ in \unit{\sec}, yielding a sampling rate $f_{\mathrm{s}} = \frac{1}{k}$, and define a vector time series $\bm{q}^n = \bm{q}(nk),\;n=0,\dots,N-1$ on a uniform time grid.

To arrive at a one-step update, we rewrite \eqref{eq: modal_equation} as a first-order system by introducing modal velocities $\bm{p} = [p_1(t),\dots,p_M(t)]\tran$:
\[
\begin{cases}
\dot{\bm{q}}
=
\bm{p} 
\\
\dot{\bm{p}}
=
-
2 \bm{\mathrm{S}}
\bm{p}
-
\bm{\Omega}^2
\bm{q}
+
\gamma^2
\bm{f}(\bm{q})
+
\bm{\Phi}(x_{\mathrm{e}})
f_{\mathrm{e}}(t)
\end{cases}
\label{eq: modal_equation_first}
\numberthis
\]
Using centred approximation for time derivatives, we get:
\[
\begin{cases}
\frac{\bm{q}^{n+1} - \bm{q}^n}{k}
=
\bm{p}^{n+\frac{1}{2}}
\\[0.5em]
\frac{
\bm{p}^{n+\frac{1}{2}}
-
\bm{p}^{n-\frac{1}{2}}
}{k}
=
-
2 \bm{\mathrm{S}}
\bm{p}^n
-
\bm{\Omega}^2
\bm{q}^n
+
\gamma^2
\bm{f}(\bm{q}^n)
+
\bm{\Phi}(x_{\mathrm{e}})
f_{\mathrm{e}}^n
\end{cases}
\label{eq: verlet_interleaved}
%\numberthis
\]
where $\bm{p}^{n+\frac{1}{2}}$ is an interleaved sequence of modal velocities and $f_{\mathrm{e}}^n = f_{\mathrm{e}}(nk)$.

Following \cite{Hairer2003}, this numerical scheme can be written in an explicit form to produce an update $(\bm{q}^n, \bm{p}^n) \rightarrow (\bm{q}^{n+1}, \bm{p}^{n+1})$:
\[
\begin{cases}
\bm{p}^{n+\frac{1}{2}}
=
\bm{p}^n
+
\frac{k}{2}
\big[
-
2 \bm{\mathrm{S}}
\bm{p}^n
-
\bm{\Omega}^2
\bm{q}^n
+
\gamma^2
\bm{f}(\bm{q}^n)
+
\bm{\Phi}(x_{\mathrm{e}})
f_{\mathrm{e}}^n
\big]
\\[0.5em]
\bm{q}^{n+1}
=
\bm{q}^n
+
k
\bm{p}^{n+\frac{1}{2}}
\\[0.5em]
\bm{p}^{n+1}
=
\big(
\bm{\mathrm{I}}
+
k
\bm{\mathrm{S}}
\big)^{-1}
\big[
\bm{p}^{n+\frac{1}{2}}
+{}
\\[0.5em]
\phantom{\bm{p}^{n+1}}
{}+
\frac{k}{2}
\big(
-
\bm{\Omega}^2
\bm{q}^{n+1}
+
\gamma^2
\bm{f}(\bm{q}^{n+1})
+
\bm{\Phi}(x_{\mathrm{e}})
f_{\mathrm{e}}^{n+1}
\big)
\big]
\end{cases}
\label{eq: verlet}
\numberthis
\]
Since matrix $\bm{\mathrm{I}}+k\bm{\mathrm{S}}$ is diagonal, its inverse can be easily computed. Using \eqref{eq: superposition}, we obtain an audio output as $w^n = \bm{\Phi}\tran(x_{\mathrm{o}}) \bm{q}^n$.

%%%%%%%%%% LEARNING ALGORITHM %%%%%%%%%%
\section{Learning Algorithm}
\label{sec: algorithm}

%%%%%%%%%% NEURAL ODES %%%%%%%%%%
\subsection{Neural Ordinary Differential Equations (NODEs)}

NODEs can be defined through the following first-order system:
\[
\frac{d\bm{y}}{dt}
=
\bm{g}_{\theta}
(\bm{y}, t)
,
\quad
\bm{y}(0)
=
\bm{y}_0.
\label{eq: neural_ode}
\numberthis
\]
Here $\bm{y} = \bm{y}(t)\!\!: \mathbb{R}^+ \rightarrow \mathbb{R}^K$ is an unknown function of time $t$, $\bm{y}_0 \in \mathbb{R}^K$ is an initial condition and $\bm{g}_{\theta}(\bm{y}, t)\!\!: \mathbb{R}^K \times \mathbb{R}^+ \rightarrow \mathbb{R}^K$ is a neural network where $\theta$ denotes the set of all learnable parameters and $K$ denotes the state dimension. Generally, a simple neural architecture such as an MLP or convolutional network is chosen for $\bm{g}_{\theta}(\bm{y}, t)$. Chen et al.\@ \cite{Chen2018} have showed that the system \eqref{eq: neural_ode} in combination with a numerical solver, labelled as ODE-Net, can be trained from data to reproduce dynamics of a target system.

Assume a target trajectory $\tau = \{\bm{y}_0,\bm{y}^1,\dots,\bm{y}^{N-1}\}$ sampled on a uniform time grid with a step $k$. Given a predicted trajectory $\widetilde{\tau} = \{\bm{y}_0,\widetilde{\bm{y}}^1,\dots,\allowbreak\widetilde{\bm{y}}^{N-1}\}$ by a numerical solution to the initial value problem \eqref{eq: neural_ode}, i.e., a forward pass of the ODE-Net, we can construct an objective function $\mathcal{J}(\theta)$ such as a mean squared error (MSE):
\[
\mathcal{J}(\theta)
=
\frac{1}{KN}
\sum_{n=0}^{N-1}
\norm{
\widetilde{\bm{y}}^n
-
\bm{y}^n
}_2^2
,
\label{eq: mse}
\numberthis
\]
where $\norm{\cdot}_2$ is the Euclidean norm. We search for a local minimum of $\mathcal{J}(\theta)$ using gradient-based optimisation techniques where the gradient $\nabla_{\theta}\mathcal{J}$ can be computed using the backpropagation algorithm \cite{Rumelhart1986} through internal operations of a numerical solver or the adjoint sensitivity method \cite{Pontryagin1962, Chen2018}. In most cases the objective function $\mathcal{J}(\theta)$ will be averaged for a finite set of target trajectories before each optimisation step.

%%%%%%%%%% EXTENSION FOR PHYSICAL MODELLING SYNTHESIS %%%%%%%%%%
\subsection{Extension for Physical Modelling Synthesis}

In the case of modal synthesis, there is a known ODE structure \eqref{eq: modal_equation_first} which can serve an inductive bias for a NODEs framework \cite{KidgerPhD}. In particular, we parametrise only a dimensionless memoryless nonlinear function $\bm{f}_{\theta}(\bm{q})\!\!: \mathbb{R}^M \rightarrow \mathbb{R}^M$ with a neural network, yielding a system of physically-informed NODEs:
\[
\begin{bmatrix}
\dot{\bm{q}} \\
\dot{\bm{p}}
\end{bmatrix}
=
\underbrace{
\begin{bmatrix}
\hphantom{-}
\bm{0}
\hphantom{^2}
&
\hphantom{-}
\bm{\mathrm{I}}
\hphantom{\bm{\mathrm{S}}}
\\
-\bm{\Omega}^2
&
-2\bm{\mathrm{S}}
\end{bmatrix}
\begin{bmatrix}
\bm{q} \\
\bm{p}
\end{bmatrix}
}_{\mathclap{\text{Linear vibration}}}
{}+{}
\gamma^2
\underbrace{
\begin{bmatrix}
\bf{0} \\
\bm{f}_{\theta}(\bm{q})
\end{bmatrix}
}_{\mathclap{\text{Neural network}}}
+
\underbrace{
\begin{bmatrix}
\bf{0} \\
\bm{\Phi}(x_{\mathrm{e}})
\end{bmatrix}
f_{\mathrm{e}}(t)
}_{\mathclap{\text{Excitation}}}
\label{eq: modal_equation_first_neural}
\numberthis
\]
As mentioned earlier, initial conditions for a state vector $\bm{y}\tran = [\bm{q}\tran, \bm{p}\tran]$ are assumed to be zero. To compute a forward pass of the physically-informed ODE-Net, we:
\begin{itemize}
    \item set $\bm{\mathrm{S}}, \bm{\Omega}, \gamma, \bm{\Phi}(x_{\mathrm{e}})$ in \eqref{eq: modal_equation_first_neural} using physical parameters of a target solution;
    \item precompute $f_{\mathrm{e}}^n = f_{\mathrm{e}}(nk),\;n=0,\dots,N-1$ using \eqref{eq: excitation} and excitation parameters $f_{\mathrm{amp}}, T_{\mathrm{e}}$ of a target solution;
    \item use the St{\"o}rmer-Verlet method for \eqref{eq: modal_equation_first_neural} as in the case of a regular system \eqref{eq: verlet} to produce a predicted trajectory $\widetilde{\tau}$.
\end{itemize}

The formulation \eqref{eq: modal_equation_first_neural} has strong implications. First, the exact expression for linear vibration exploits the periodic, harmonic and lossy nature of a musical system. Thus, we aid optimisation of the network by constraining the space of possible solutions and improve interpretability of the model.

Second, the neural network $\bm{f}_{\theta}(\bm{q})$ is memoryless and dimensionless, thus does not depend on physical parameters of a system and external excitation. Theoretically, these parameters can be changed after the training as long as range of modal displacements $\bm{q}$ stays the same as in a training dataset to simulate other configurations of a system.

Finally, we are able to use numerical methods developed directly for second-order systems (e.g., the St{\"o}rmer-Verlet method) that are commonly used in musical acoustics. In the linear case, some such methods allow for exact discretisation of the harmonic oscillator equation \cite{BilbaoNSS}.

%%%%%%%%%% EVALUATION %%%%%%%%%%
\section{Evaluation}
\label{sec: evaluation}

The physically-informed ODE-Net described in Section \ref{sec: algorithm} has been implemented in the PyTorch framework \cite{Paszke2019} and evaluated for two case studies: nonlinear oscillator (Section \ref{sec: nonlinear_oscillator_evaluation}) and nonlinear transverse string (Section \ref{sec: nonlinear_string_evaluation}). The training was conducted on cloud servers equipped with NVIDIA GeForce RTX 2080 Ti GPUs. The source code used for dataset generation and training is available in the accompanying GitHub repository\footnote{\url{https://github.com/victorzheleznov/dafx25}}.

For the parametrisation of $\bm{f}_{\theta}(\bm{q})$ we use an MLP with a linear output layer and a varying number of hidden layers of $100$ units. Leaky rectified linear units (Leaky ReLUs) are used as activation functions in hidden layers and Kaiming initialisation \cite{He2015} is used for the initial weights of the network. It was noted that saturating activation functions such as the hyperbolic tangent function might cause a vanishing gradient problem when modal displacements $\bm{q}$ span a wide range. Moreover, we are interested in reducing the computational cost of $\bm{f}_{\theta}(\bm{q})$ compared to the target nonlinear function $\bm{f}(\bm{q})$ \eqref{eq: nonlinearity_modal}, making rectified linear units a compelling choice. Compared to regular ReLUs, Leaky ReLUs provided faster convergence in optimisation in our experiments.

The training loss is the MSE taken over the whole state vector \eqref{eq: mse}, i.e., including both modal displacements $\bm{q}$ and modal velocities $\bm{p}$. Backpropagation is performed using internal operations of the numerical solver \eqref{eq: verlet}, i.e., the "discretise-then-optimise" method, which is generally a preferred approach due to its gradient accuracy, speed and straightforward implementation \cite{KidgerPhD}. The Adam optimiser \cite{Kingma2017} is used with default parameters. The dataset is divided into two subsets: $80\%$ is used for training and $20\%$ is used for validation. The training is performed for $5000$ epochs. The resulting model is chosen based upon the lowest loss obtained on the validation subset.

For training we use a variation of the teacher forcing technique \cite{GoodfellowDL} by splitting up a target trajectory into $1$ \unit{\milli\second} segments and providing true initial conditions for each segment to the ODE-Net. Since the numerical method \eqref{eq: verlet} is given as a one step update we have access to both displacement and velocity of the target numerical solution at each time step, and thus initial conditions for each segment. In addition, the excitation function \eqref{eq: excitation} is shifted in time to reflect a new starting point of integration. The main reason for using this technique is to speed up training as the number of integration steps in the numerical solver is significantly reduced. These integration steps can not be parallelised in time. Moreover, the likelihood of vanishing and exploding gradient problems during optimisation is also reduced by this technique as we avoid backpropagation on long time series \cite{KidgerPhD}.

%%%%%%%%%% NONLINEAR OSCILLATOR %%%%%%%%%%
\subsection{Nonlinear Oscillator}
\label{sec: nonlinear_oscillator_evaluation}

Up to this point, discussion has been centred around the nonlinear transverse string model. To illustrate the generality of the proposed approach, consider a simple nonlinear oscillator of the following form:
\[
\ddot{q}
+
\omega_0^2
q
=
\gamma^2
f(q)
+
f_{\mathrm{e}}(t),
\label{eq: oscillator}
\numberthis
\]
where $\omega_0 = 400$ and $\gamma = 110$.

We generate two datasets consisting of $60$ one-second trajectories at $44.1$ \unit{\kilo\hertz} using exactly the same set of randomly-generated external excitations. The first dataset is created using the cubic nonlinearity $f(q) = -q^3$ and the other using the hyperbolic sine nonlinearity $f(q) = -\sinh(q)$. We use an MLP with two hidden layers of $100$ units to parametrise $f_{\theta}(q)$ to ensure that the network has enough capacity to learn the underlying nonlinearities.

\begin{figure}[ht]
\center
\includegraphics[width=\columnwidth]{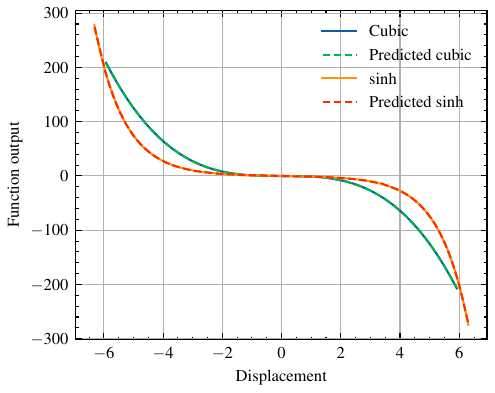}
\caption{{\it Target and predicted nonlinear functions for the oscillator \eqref{eq: oscillator}.}}
\label{fig: nonlinearity}
\end{figure}

This lumped case is especially useful for analysis as we are able to easily visualise the learned nonlinear functions. As seen in fig.\@ \ref{fig: nonlinearity}, the network is capable of reproducing both the cubic and the hyperbolic sine nonlinearities and can distinguish between them from data. Ranges of displacement for both cases correspond to minimum and maximum values in the datasets.

%%%%%%%%%% NONLINEAR TRANSVERSE STRING %%%%%%%%%%
\subsection{Nonlinear Transverse String}
\label{sec: nonlinear_string_evaluation}

For the case of nonlinear transverse string, which is described by the multi-dimensional system \eqref{eq: modal_equation}, we need to choose the number of modes we will use in simulations. In this work we have settled on $100$ modes. Taking into account the effect of stiffness, this covers most of the audible range for the chosen string parameters --- up to $17$ \unit{\kilo\hertz} for the lowest considered fundamental frequency at around $60$ Hz. Simulation of this wide range of frequencies requires oversampling by two for standard sampling rates such as $44.1$ \unit{\kilo\hertz} and $48$ \unit{\kilo\hertz} to avoid instability of the St{\"o}rmer-Verlet method and aliasing due to the nonlinear effects for high amplitude excitations.

%%%%%%%%%% DATASETS %%%%%%%%%%
\subsubsection{Datasets}

Simulation parameters used for the generation of two datasets --- one for training and validation and the other for testing --- are outlined in table \ref{tab: dataset_parameters}, where $T_{\mathrm{sim}}$ corresponds to the duration of the simulation. Each dataset consists of $60$ string trajectories, which include both displacement and velocity information for each mode. For training and validation, only one set of physical parameters is chosen corresponding to a $61.72$ \unit{\hertz} string. This string is excited by randomly-generated excitation functions \eqref{eq: excitation} at randomised excitation positions $x_{\mathrm{e}}$ using a uniform distribution for the specified parameter ranges. For testing, string parameters $\gamma$ and $\kappa$ are also randomly generated to produce a range of fundamental frequencies from around $65$ \unit{\hertz} to $123$ \unit{\hertz}. The simulation duration is increased to account for a longer decay time due to a smaller damping parameter $\sigma_0$. In addition, the sampling rate is increased from $88.2$ \unit{\kilo\hertz} to $96$ \unit{\kilo\hertz}. Even though the ranges of the excitation parameters are the same for the two datasets, excitation functions are generated independently. For both datasets, audio output is drawn from randomised positions $x_{\mathrm{o}}$ along a string for each trajectory.

These parameters are motivated by two considerations. First, we want to test generalisation of the model to physical parameters, sampling rates and time scales not seen during training. In view of other machine learning approaches, this flexibility and controllability of the physically-informed ODE-Net can be considered as its main advantage. Second, strings with low fundamental frequencies are chosen so that the nonlinear effect is more prominent in simulations \cite{BilbaoNSS}. Since the network architecture is designed to learn the residual between the linear and nonlinear solutions, the training and validation dataset needs to reflect a significant difference between them.

\begin{table}[ht]
\def\arraystretch{1.3}
\centering
\caption{\it Simulation parameters used for dataset generation.}
\label{tab: dataset_parameters}
\begin{tabular}{|l|c|c|}
\hline
\textbf{Parameter} & \begin{tabular}{@{}c@{}}\textbf{Training and}\\\textbf{Validation}\end{tabular} & \textbf{Test} \\
\hline\hline
$f_{\mathrm{s}}$
&
$88.2$ \unit{\kilo\Hz}
&
$96$ \unit{\kilo\Hz} \\\hline
$T_{\mathrm{sim}}$
&
$2$ \unit{\sec}
&
$3$ \unit{\sec}
\\\hline
$T_{\mathrm{e}}$
& 
$[0.5, 1.5]$ \unit{\milli\second}
& 
$[0.5, 1.5]$ \unit{\milli\second} 
\\\hline
\hline
$\gamma$ & $123.4$ & $[130, 246]$ \\\hline
$\kappa$ & $1.01$ & $[1.01, 1.1]$ \\\hline
$\sigma_0$ & $3$ & $2$ \\\hline
$\sigma_1$ & $\num{2e-4}$ & $\num{2e-4}$ \\\hline
$x_{\mathrm{e}}$ & $[0.1, 0.9]$ & $[0.1, 0.9]$ \\\hline
$x_{\mathrm{o}}$ & $[0.1, 0.9]$ & $[0.1, 0.9]$ \\\hline
$f_{\mathrm{amp}}$
& 
$[2, 3] \times 10^4$
& 
$[2, 3] \times 10^4$
\\
\hline
\end{tabular}
\end{table}

%%%%%%%%%% RESULTS %%%%%%%%%%
\subsubsection{Results}

For training we use an MLP with five hidden layers of $100$ units to parametrise $\bm{f}_{\theta}(\bm{q})$. Although no extensive research was conducted on the optimal size of the network with regards to simulation accuracy and computational cost, this structure provided the best empirical results when training MLPs with varying number of hidden layers. Further optimisation beyond $5000$ epochs did not provide significant reduction in the validation loss, suggesting that the model has converged to a local minimum.

For evaluation, we are using the relative MSE for displacement trajectory $\bm{q}^n$ and audio output $w^n$:
\[
\frac{
\sum_n
\norm{\widetilde{\bm{q}}^n - \bm{q}^n}_2^2
}
{
\sum_n
\norm{\bm{q}^n}_2^2
}
,
\quad
\frac{
\sum_n
\module{\widetilde{w}^n - w^n}^2
}
{
\sum_n
\module{w^n}^2
},
\]
where $\widetilde{\bm{q}}^n$ and $\widetilde{w}^n$ are predictions given by the physically-informed ODE-Net. These metrics are chosen as the audio output of simulation is directly dependent on the displacement trajectory of a string.

Metrics for the training and validation dataset and the test dataset are provided in table \ref{tab: metrics}. Metrics are evaluated for the initial $100$ \unit{\milli\second} and for the full duration of simulation, i.e., $2$ or $3$ \unit{\sec}, respectively. As can be seen, metrics for the test dataset remain on the same order of magnitude compared to the training and validation dataset, especially for the initial $100$ \unit{\milli\second}. This suggests that the performance of the trained physically-informed ODE-Net does not significantly degrade for unseen simulation parameters. 

\begin{table}[ht]
\def\arraystretch{1.3}
\sisetup{detect-all = true}
\centering
\caption{\it Metrics for the nonlinear transverse string datasets.}
\label{tab: metrics}
\begin{tabular}{|l|c|c|}
\hline
\textbf{Metric} & \begin{tabular}{@{}c@{}}\textbf{Training and}\\\textbf{Validation}\end{tabular} & \textbf{Test} \\
\hline\hline
\multicolumn{3}{|c|}{
\it
Evaluated for initial
100 \unit{\milli\second}
} \\
\hline\hline
Rel. MSE for displacement
&
$\num{3.91e-03}$
&
$\num{5.07e-03}$
\\\hline
Rel. MSE for output
&
$\num{3.66e-03}$
&
$\num{5.16e-03}$
\\
\hline\hline
\multicolumn{3}{|c|}{
\it
Evaluated for full duration
} \\\hline
\hline
Rel. MSE for displacement
&
$\num{3.64e-02}$
&
$\num{6.68e-02}$
\\\hline
Rel. MSE for output
&
$\num{3.54e-02}$
&
$\num{7.00e-02}$
\\
\hline
\end{tabular}
\sisetup{detect-none = true}
\end{table}

\begin{figure}[ht]
\center
\includegraphics[width=\columnwidth]{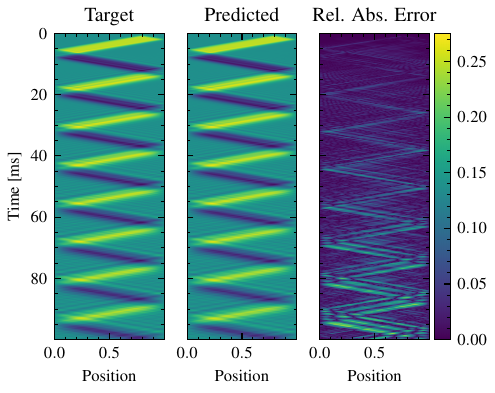}
\caption{{\it Displacement trajectory for the selected test string. On the right, the relative absolute error between the target and predicted trajectories is shown, normalised by the maximum absolute value of target trajectory.}}
\label{fig: 9_displacement_grid}
\end{figure}

\begin{figure*}[p]
\center
\includegraphics[width=\textwidth]{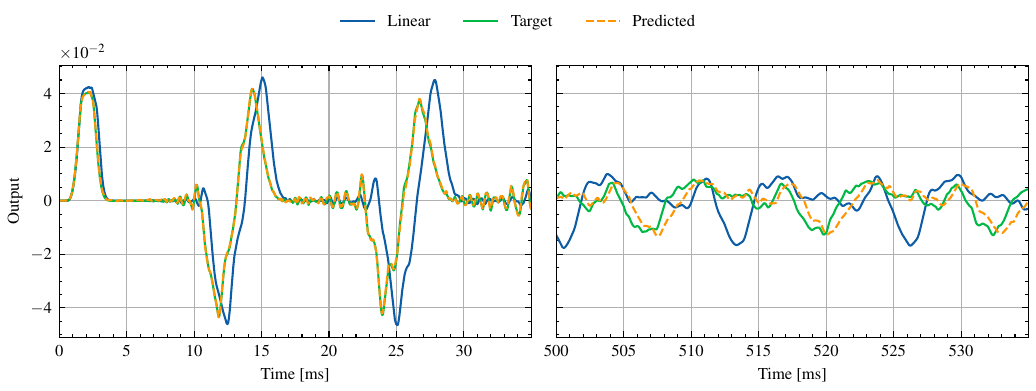}
\caption{{\it Output for the selected test string at normalised position $\mathit{x_o}$ equal to 0.87}.}
\label{fig: 9_wave}
\end{figure*}

\begin{figure*}[p]
\center
\includegraphics[width=\textwidth]{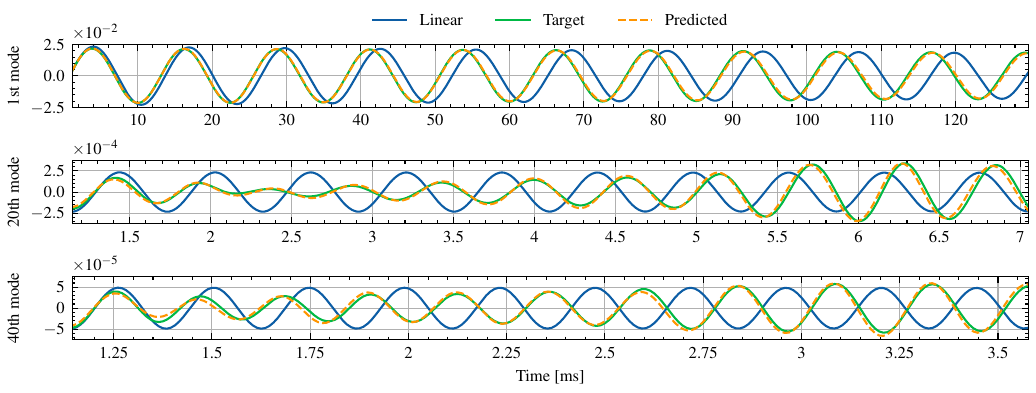}
\caption{{\it Displacements of the 1st, 20th, 40th mode for the selected test string at normalised position $\mathit{x_o}$ equal to 0.87 (initial 10 periods).}}
\label{fig: 9_displacement}
\end{figure*}

To illustrate a specific example, we select a trajectory from the test dataset with the largest relative MSE for audio output considering the full simulation duration. This corresponds to a $77.72$ \unit{\hertz} string. Fig.\@ \ref{fig: 9_displacement_grid} shows the displacement trajectory of the selected test string for the initial $100$ \unit{\milli\second}. As can be seen, the predicted trajectory maintains the structure of target solution but starts to lag behind with time. This is to be expected, as any difference between the approximated and the underlying nonlinearity will be integrated over time by a numerical solver, thus accumulating the error. This is also confirmed by the metrics in table \ref{tab: metrics} which rise when evaluated for the full duration of simulation compared to the initial $100$ \unit{\milli\second}. Examining the output waveform in fig.\@ \ref{fig: 9_wave}, we see that initially the predicted waveform closely follows the target solution, including high-frequency partials of higher modes. Moving forward in time, the predicted waveform still resembles the target solution much closer compared to the linear solution. Looking at displacements for individual modes in fig.\@ \ref{fig: 9_displacement}, we also see that initially the predicted displacements follow the target solution. It should be noted that as vibration amplitude decreases over time due to loss in the system the nonlinear effects become less prominent \cite{BilbaoNSS}, thus the initial response of the model to an external excitation is significantly more important for capturing the nonlinear behaviour.

\begin{figure*}[p]
\sisetup{detect-all = true}
\center
\includegraphics[width=\textwidth]{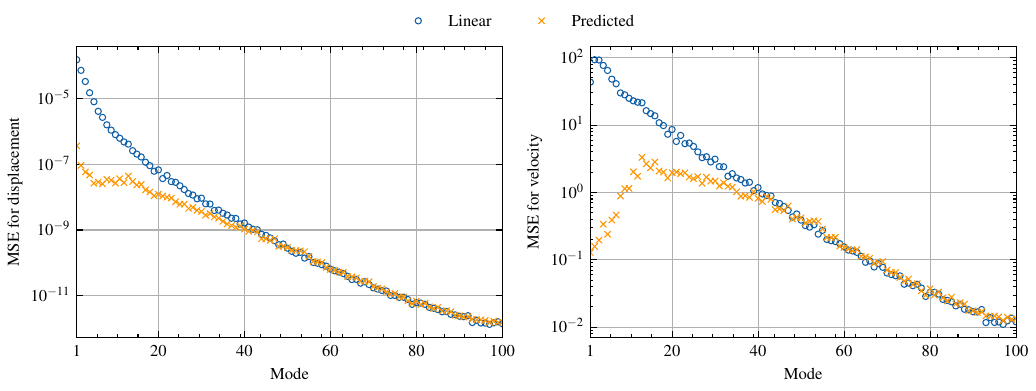}
\caption{{\it MSE per mode for the initial 100 \unit{\milli\second} of the linear and predicted string trajectories for the test dataset.}}
\label{fig: mse_per_mode_slice_test}
\sisetup{detect-none = true}
\end{figure*}

\begin{figure*}[ht]
\center
\includegraphics[width=\textwidth]{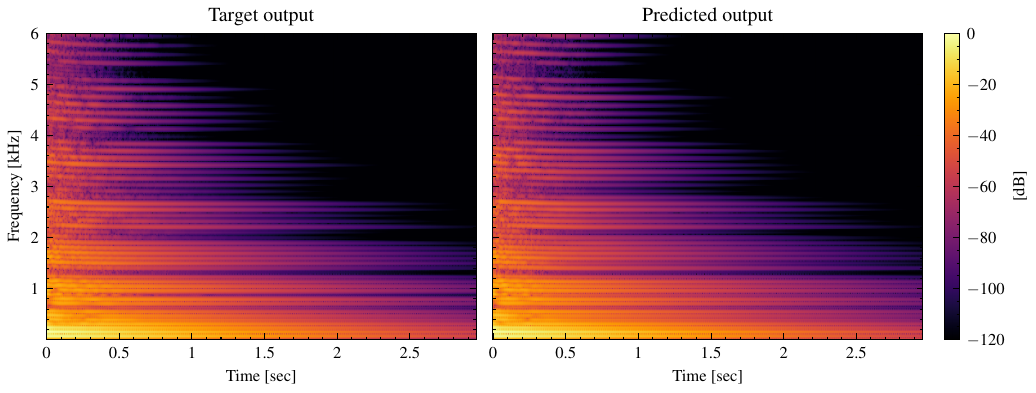}
\caption{{\it Output spectrogram for the selected test string at normalised position $\mathit{x_o}$ equal to 0.87.}}
\label{fig: 9_spec}
\end{figure*}

We can examine the model further by comparing it to the linear solution. Fig.\@ \ref{fig: mse_per_mode_slice_test} shows MSE for the initial $100$ \unit{\milli\second} of the linear and predicted trajectories compared to the target solution individually for each mode, evaluated over the whole test dataset. It can be clearly seen that the predicted trajectories capture displacements and velocities up to around $40$th mode much more accurately compared to the linear solution. Since absolute values of MSE --- the objective function of optimisation --- are much lower for higher modes, the network seems to struggle to reproduce them. In a case by case examination of modes (e.g., the $40$th mode in fig.\@ \ref{fig: 9_displacement}), it was noted that the network is capable of accurately capturing a first few \unit{\milli\second} of the target solution. Afterwards, the main cause for the error of predicted trajectories becomes incorrectly estimated amplitude rather than instantaneous frequency. As can be seen on the output spectrogram for the selected test string (fig.\@ \ref{fig: 9_spec}), for which the $40$th mode corresponds to around $4$ \unit{\kilo\hertz}, the pitch glide effect is reproduced for higher modes in line with the target solution.

Considering the computational cost in this particular example, computation of $\bm{f}_{\theta}(\bm{q})$ requires far fewer floating point operations compared to the underlying nonlinearity $\bm{f}(\bm{q})$ \eqref{eq: nonlinearity_modal}. For $100$ modes, the tensor $A_{i_1,i_2,i_3}^m$ \eqref{eq: tensor} consists of $N_A = 2597200$ non-zero elements when accounted for the symmetries. Thus, it will require $\mathcal{O}(N_A)$ floating point operations to compute $\bm{f}(\bm{q})$. On the other hand, $\bm{f}_{\theta}(\bm{q})$ consists of five hidden layers of $100$ units with Leaky ReLUs and a linear output layer. Computation of $\bm{f}_{\theta}(\bm{q})$ takes $121000$ summations and multiplications combined when using a naive matrix multiplication algorithm. This suggests that the presented hybrid approach with a relatively small neural network might be able to sufficiently capture the behaviour of a complex physical model, while reducing the computational cost compared to the regular modal synthesis method. As of now, there were no formal listening tests conducted to assess a trade-off in perceptual accuracy, but readers are encouraged to listen to audio examples presented on the accompanying page\footnote{\url{https://victorzheleznov.github.io/dafx25}}.

%%%%%%%%%% CONCLUSIONS %%%%%%%%%%
\section{Conclusions}
\label{sec: conclusions}

A method for modelling distributed musical systems in a modal form using neural ordinary differential equations has been considered here. The proposed approach separates the problem into the linear and nonlinear parts, trying to combine complementary strengths of modal synthesis and machine learning. The analytical solution for linear vibration of modes allows the isolation of physical system parameters and external excitation from the neural network, which learns a dimensionless and memoryless nonlinear coupling between the modes. It has been shown that such structure can be used for simulation of a system with physical parameters unseen during training without significant degradation in accuracy of the resulting solution. In addition, the use of a neural network allows for an efficient representation of a complex nonlinear function for the non-trivial case study of nonlinear transverse string vibration.

This paper leaves many avenues and unanswered questions for further work, several of which will be explored in the near future. Currently, the model uses both displacement and velocity information of each mode for learning. However, as modal displacements and velocities are inherently connected in a system of ordinary differential equations, it should be possible for the model to learn using just the displacement information or one-dimensional audio output taken at randomised positions along the string. This would simplify a comparison of the proposed approach to other machine learning methods \cite{Parker2022, Diaz2024, Lee2024}. For such a comparison, external forcing terms should be omitted.

The St{\"o}rmer-Verlet method \cite{Hairer2003} used for simulation here is an efficient numerical method that allows for an explicit update equation. Nevertheless, even for strings with low fundamental frequencies, it requires oversampling to run in a stable manner. An extension to the scalar auxiliary variable technique \cite{Bilbao2023, Russo2024} seems natural as it is the only numerical method known to the authors which is not only explicit but both conditionally stable and general, i.e., can be applied to different nonlinearities using the same discretisation process.

Further research is required into potential computational savings to be had with a neural network compared to the underlying nonlinearity, as real-time capability of the presented model remains unexplored. There should be a crossing point in terms of number of modes and complexity of the network where the network becomes more efficient than the analytical expression. With these computational savings, a trade-off in perceptual accuracy of the resulting sound should be studied in formal listening tests.

Finally, leaving the realm of computer simulation, the presented model has a potential for identification of nonlinear acoustic systems by learning from real-world data. Bridging the gap between the theory and what is observed in practice using such hybrid approaches has already been suggested in other fields \cite{KidgerPhD}. Considering musical acoustics, there are cases such as the bowed string where the underlying friction characteristic between the string and the
bow is yet to be fully understood \cite{Galluzzo2017}.

\bibliographystyle{IEEEbib}
\begin{footnotesize}
\bibliography{DAFx25_tmpl}% requires file DAFx25_tmpl.bib
\end{footnotesize}

\end{document}